\pdfoutput=1
\documentclass[lettersize,journal]{IEEEtran}
\usepackage{amsmath,amsfonts}
\usepackage{algorithmic}
\usepackage{array}
\usepackage[caption=false,font=normalsize,labelfont=sf,textfont=sf]{subfig}
\usepackage{textcomp}
\usepackage{stfloats}
\usepackage{url}
\usepackage{verbatim}
\usepackage{graphicx}
\def\BibTeX{{\rm B\kern-.05em{\sc i\kern-.025em b}\kern-.08em
    T\kern-.1667em\lower.7ex\hbox{E}\kern-.125emX}}
\usepackage{balance}
\begin{document}
\title{Non-control-Data Attacks and Defenses:\\  A review}
\author{Lei Chong
\thanks{}}

\markboth{Journal of \LaTeX\ Class Files}%
{How to Use the IEEEtran \LaTeX \ Templates}

\maketitle

\begin{abstract}
In recent years, non-control-data attacks have become a research hotspot in the field of network security, driven by the increasing number of defense methods against control-flow hijacking attacks. These attacks exploit memory vulnerabilities to modify non-control data within a program, thereby altering its behavior without compromising control-flow integrity. Research has shown that non-control-data attacks can be just as damaging as control-flow hijacking attacks and are even Turing complete, making them a serious security threat. However, despite being discovered long ago, the threat of non-control-data attacks has not been adequately addressed. In this review, we first classify non-control-data attacks into two categories based on their evolution: security-sensitive function attacks and data-oriented programming (DOP) attacks. Subsequently, based on the non-control-data attack model, we categorize existing defense methods into three main strategies: memory safety, data confidentiality, and data integrity protection. We then analyze recent defense techniques specifically designed for DOP attacks. Finally, we identify the key challenges hindering the widespread adoption of defenses against non-control-data attacks and explore future research directions in this field.
\end{abstract}


\section{Introduction}
\IEEEPARstart{M}{emory}corruption attacks are one of the most common ways to compromise computer systems. Since C and C++, as system-level programming languages, are widely used for developing operating system kernels and browsers, they provide programmers with greater control over code, enabling them to write optimized and efficient programs. However, this also inevitably leads to security vulnerabilities, such as buffer overflows, use-after-free (UAF), and other forms of memory-related flaws.
In real-world environments, no effective defense software has been deployed for memory safety, as countermeasures ensuring complete memory safety often incur significant runtime performance overhead. For example, SoftBound+CETS imposes an average performance overhead of 116\%~\cite{nagarakatte2010cets}. As a result, the battle for memory safety continues. Attackers exploit these memory vulnerabilities to carry out memory corruption attacks in pursuit of their malicious objectives.Memory corruption attacks can be classified into three major categories \cite{szekeres2013sok} : i) control flow hijacking attacks, ii) non-control-data attacks, iii) information leakage attacks.

Control Flow Hijacking Attacks are a very common attack method in which attackers hijack control flow by modifying control data in memory (e.g., return addresses or code pointers). These attacks include code injection attacks \cite{one1996smashing}, Return-into-libc attacks \cite{designer1997bugtraq}, and Return-Oriented Programming (ROP) attacks \cite{shacham2007geometry,checkoway2010return,bletsch2011jump,bosman2014framing,schuster2015counterfeit}.
Code injection attacks hijack control flow by controlling the thread to call the return address stored in the stack or the jump target stored in a register. To defend against such attacks, researchers have proposed various countermeasures, such as StackGuard \cite{cowan1998stackguard} and Data Execution Prevention (DEP) \cite{andersen2004changes}. Particularly, with the introduction of DEP technology, shellcodes injected directly into memory by attackers are no longer executable due to the lack of execution permissions, rendering such attacks ineffective. However, some researchers proposed utilizing existing code in memory to bypass DEP defenses, which led to Return-into-libc attacks \cite{designer1997bugtraq}.
Since Return-into-libc attacks can only reuse library function binaries, their attack capabilities are limited. To overcome this limitation, Shacham proposed ROP attacks in 2007 and proved that they are Turing-complete \cite{shacham2007geometry}. The core idea of ROP attacks is to execute malicious operations using gadgets—small code snippets that end with RET instructions—found in the target program's memory, rather than calling entire functions.
With the continuous evolution of attack and defense technologies, ROP attacks have led to various new attack variants, such as Jump-Oriented Programming (JOP) \cite{checkoway2010return,bletsch2011jump}, Sigreturn-Oriented Programming (SROP) \cite{bosman2014framing}, and Counterfeit Object-Oriented Programming (COOP) \cite{schuster2015counterfeit}. To resist these types of attacks, researchers have proposed several defense mechanisms, including Address Space Layout Randomization (ASLR) \cite{team2003pax,backes2014oxymoron}, Control-Flow Integrity (CFI) \cite{abadi2009control,zhang2013practical,mashtizadeh2015ccfi,niu2014modular,ding2017efficient}, and Code-Pointer Integrity (CPI) \cite{kuznetzov2018code}.
With the deployment of these defense mechanisms, control-flow hijacking attacks have become increasingly difficult to execute. As a result, researchers have shifted their focus toward non-control-data attacks \cite{chen2005non,hu2015automatic,carlini2015control,hu2016data,ispoglou2018block}.

Non-control-data attacks compromise computer systems by modifying the non-control data (such as user identity data, configuration data, user input data, and decision-making data) without altering the program's control flow. These attacks are as harmful as control-flow hijacking attacks and have even been proven Turing-complete. Such attacks can bypass DEP (Data Execution Prevention), ASLR (Address Space Layout Randomization), and CFI (Control-Flow Integrity), which have been commercially deployed as defense mechanisms. Although researchers have proposed numerous defense schemes against non-control-data attacks, these schemes are difficult to deploy in real-world environments due to the following three reasons: excessive runtime overhead, which negatively impacts system performance; incompatibility with existing software or hardware, meaning that deploying the defense requires modifications to the current software and/or hardware architecture; and lack of robustness, as these defenses are often limited to protecting against only one or two specific attack types and can be easily bypassed. With a variety of common non-control-data attack methods and proposed defense schemes against non-control-data attacks, it is difficult to see how effective and efficient the different solutions are, how they compare with each other, and their limitations.

Previous literature has conducted a survey on memory corruption attacks \cite{szekeres2013sok}, but it was completed a long time ago and does not include several recent non-control-data attacks \cite{hu2015automatic,carlini2015control,hu2016data,ispoglou2018block}. The motivation of this paper is to systematize and evaluate the state-of-the-art exploitation techniques and defense mechanisms for non-control-data attacks by developing a new general non-control-data attack model, which serves as a supplement to \cite{szekeres2013sok}. This paper does not cover all solutions for non-control-data attacks but focuses on the main approaches. The contributions of this paper are as follows:
\begin{itemize}
	\item{We systematically analyze non-control-data attacks and develop a general attack model.}
	\item{Based on the attack model, we identify different security protection strategies and analyze and compare these methods.}
	\item{We examine the defense mechanisms against non-control-data attacks proposed in the past three years and highlight the current defense trends.}
	\item{By analyzing all existing defenses against non-control-data attacks, we point out why these methods are difficult to deploy commercially and suggest promising future research directions for defending against non-control-data attacks.}
\end{itemize}

The remaining parts of this paper will be divided into four sections: non-control-data attacks, defenses against non-control-data attacks, recent defenses against DOP attacks, and discussion. First, Section II discusses the evolution and classification of non-control-data attacks. Section III reviews the existing defense mechanisms against non-control-data attacks. Section IV examines recent defense strategies against DOP attacks. Finally, Section V concludes this survey.

\section{Non-Control-Data Attacks}
To address the issue of non-control-data attacks, we must first understand how various types of non-control-data attacks are executed and their similarities and differences. In this section, we discuss several representative non-control-data attacks, classify them into two categories based on their attack characteristics, and then establish a general model for non-control-data attacks. Figure 3 illustrates this attack model.

\subsection{The Evolution of Non-Control-Data Attacks}
In recent years, non-control-data attacks have attracted significant attention from the research community because they do not alter the control flow and can bypass CFI defense mechanisms. Research papers on these attacks and their countermeasures are frequently presented at security conferences. Figure 1 illustrates the evolution of non-control-data attacks from a chronological perspective. The entire evolution process can be divided into three stages:

\begin{figure}[h]
	\centering
	\includegraphics[width=3.5in]{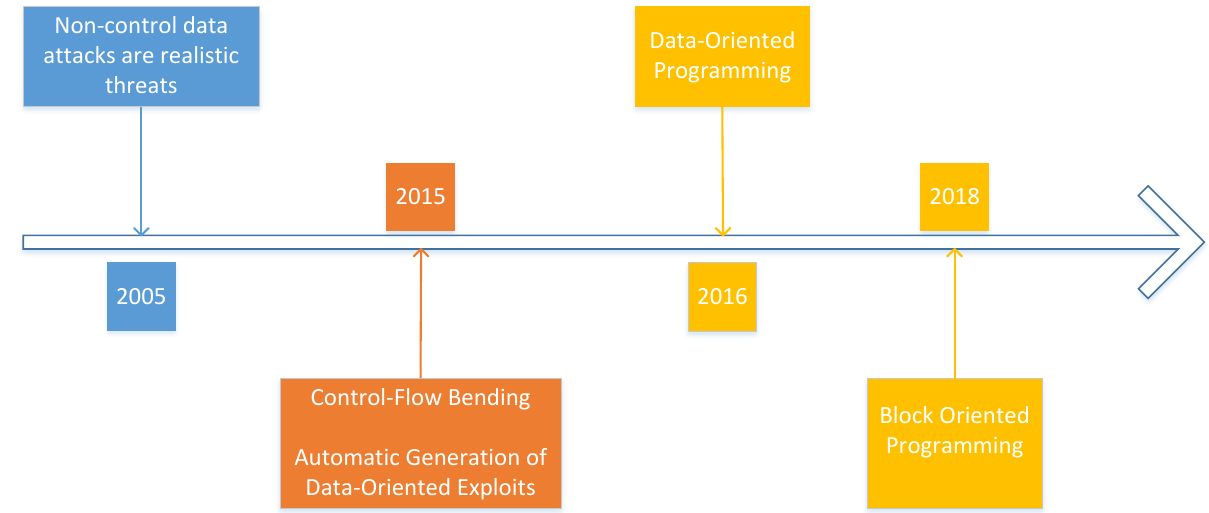}
	\caption{Evolution of non-control-data attacks.}
	\label{fig2-1}
\end{figure}

Stage 1: Chen et al. first pointed out in \cite{chen2005non} that non-control-data attacks are a real threat, as harmful as control-flow hijacking attacks. To verify this argument, they identified four types of security-critical non-control data that affect software security and constructed non-control-data attacks against these data types in several common network server applications. The results of these attacks were similar to control-flow hijacking attacks. The four security-critical data types are as follows:
1) Configuration Data: Many applications use configuration files to specify the locations of data and executable files, access control policies for files and directories, and other security- and performance-related parameters. Configuration data controls the runtime behavior of the program and is rarely changed. If an attacker corrupts this data, they can bypass the access control policy.
2) User Identity Data: Data used to identify users, such as user identifiers (UID) and group identifiers (GID). Before connecting to a remote server for authorization, the system must select appropriate permissions based on user authentication. If this value is tampered with, the attacker can gain administrator privileges.
3) User Input String: User input validation is a method used by many applications to enforce intended security policies. If an attacker uses legitimate input to pass the program’s validation check, then later alters that data and forces the program to use the modified input, they can break into the system.
4) Decision-Making Data: Boolean variables are often used in decision-making processes. If an attacker modifies these values, it can affect authentication results, enabling malicious actions such as gaining access to the target system.

The following is a concrete example to illustrate the process of a non-control-data attack. As shown in Table 1, this is an example of a vulnerable web server, wu-ftpd. The attacker exploits a format string vulnerability in the fifth line to overwrite the value of pw$\to$pw\underline{ }uid in the fourth line with the root's UID. When the program reaches the ninth line, the root privileges are not downgraded to normal user privileges as they normally would be, allowing the process to continue running with root-level privileges.

\begin{table}[h]
	\centering
	\includegraphics[width=3.5in]{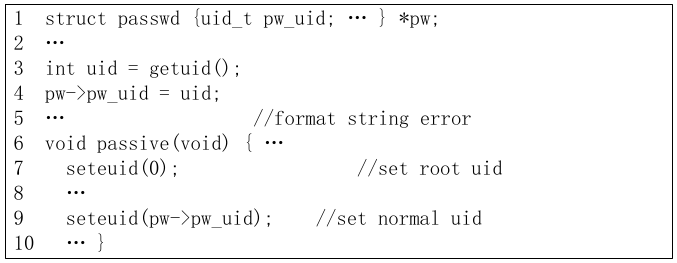}
	\caption{non-control-data attack in a vulnerable SSH server \cite{hu2015automatic}.}
	\label{table2-1}
\end{table}

Stage 2: Carlini et al. \cite{carlini2015control} and Hu et al. \cite{hu2015automatic} independently proposed general frameworks for constructing non-control-data attacks. In particular, Hu et al. introduced a novel concept known as data-flow stitching and developed a prototype system, FLOWSTITCH, designed to automate the construction of such attacks. Experimental evaluations demonstrated that all attacks generated by FLOWSTITCH successfully bypassed DEP and CFI, while more than half were also capable of circumventing ASLR, ultimately enabling information disclosure or privilege escalation.

To automate the construction of non-control-data attacks, Hu et al. proposed a concept called Two-Dimensional Data-Flow Graph (2D-DFG), which represents the data flow of a given program in two dimensions: memory address and execution time. 
A 2D-DFG is a directed graph, denoted as \( G=\{V, E\} \), where \( V \) is the set of vertices and \( E \) is the set of edges. A vertex \( v = (a,t) \) is created when the program writes to address \( a \) at execution time \( t \), where \( a \) can be either a memory address or a register name. 
In a 2D-DFG, if there exists an edge \((v', v)\) from vertex \( v' \) to vertex \( v \), it indicates a data dependency established during program execution. This edge can be classified into two types: 

\begin{itemize}
	\item Data Edge: If the value of \( v \) is derived from \( v' \), the edge is called a data edge.
	\item Address Edge: If the address of \( v \) is derived from \( v' \), the edge is called an address edge.
\end{itemize}
 
According to the example given by Hu et al., Table 1 shows a vulnerable code of wu-ftpd web server, and figure 2 shows the two 2D-DFG of this program before (left) and after (right) the attack. The attacker overwrites the value of pw$\to$pw\underline{ }uid in line 4 through the format string error in line 5 of the program, thus creating a new vertex at time 5. As the program continues to execute, a new dataflow path is created, As shown by the dashed line on the right of the figure. This example is just a single edge stitch, when a single edge stitch fails to meet the requirements, the attacker can also achieve malicious purposes by multi edge stitch composed of several single edge stitches, or pointer stitch which corrupts a value of a pointer that is later used as a pointer to a vertex in the source or target flow. 

\begin{figure}[!t]
	\centering
	\includegraphics[width=3.5in]{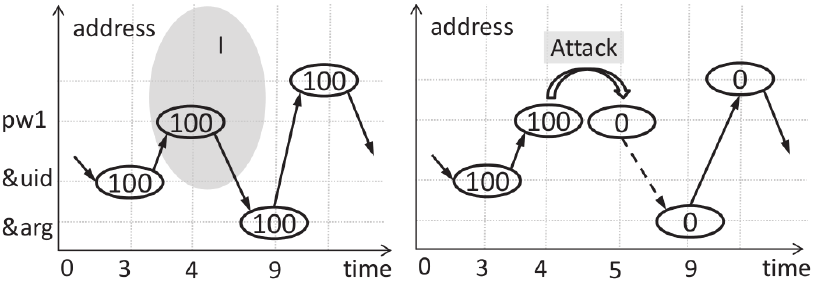}
	\caption{ Example of data-flow stitching \cite{hu2015automatic}.}
	\label{fig2-2}
\end{figure}

Carlini et al. \cite{carlini2015control} introduce Control-Flow Bending (CFB), a general framework for non-control-data attacks. This technique leverages memory errors to manipulate the parameters of security-sensitive functions, such as memcpy() and printf(), enabling malicious exploitation while remaining compliant with fine-grained Control-Flow Integrity (CFI) protections.

Stage 3, Hu et al. \cite{hu2016data} and Ispoglou et al. \cite{ispoglou2018block} demonstrated that non-control-data attacks possess Turing completeness. The BOP attack, introduced by Ispoglou et al. \cite{ispoglou2018block}, is a variant of the DOP attack proposed by Hu et al. \cite{hu2016data}. Furthermore, the framework for constructing the attack has evolved from a semi-automated approach to a fully automated one.

Hu et al. \cite{hu2016data} demonstrate that even with a single memory error, non-control-data attacks can achieve Turing-complete computations through data-oriented programming (DOP). The structure of a DOP attack resembles that of return-oriented programming (ROP) and jump-oriented programming (JOP), comprising dispatcher gadgets and data-oriented gadgets. Data-oriented gadgets are short instruction sequences within the program that simulate all specific operation instructions (e.g., arithmetic, assignment, conditional jumps) of MINDOP, a Turing-complete language. However, data-oriented gadgets differ from code gadgets in two key aspects. First, data-oriented gadgets can only use memory to pass operation results, whereas code gadgets can utilize both memory and registers. Second, data-oriented gadgets must execute within a legitimate control flow and do not need to be executed sequentially. The role of dispatcher gadgets is to link various disjoint gadgets to enable recursive computation. The dispatcher consists of a loop and a selector: the loop ensures the program continues running and executing various gadgets to fulfill the attacker’s objective, while the selector, controlled by the memory error, determines the specific data-oriented gadget to execute in each iteration.

The process of constructing a DOP attack consists of the following three steps:
1) Gadget Identification: To obtain gadgets that simulate basic MINDOP operations, Hu et al. \cite{hu2016data} statically identified the load-semantic-store chain from LLVM IR.
2) Dispatcher Identification: To find dispatchers that chain disjoint gadgets, Hu et al. \cite{hu2016data} statically identified loops containing gadgets from LLVM IR.
3) Gadget Stitching: This step was manually completed and further divided into three substeps.  
First step: For a given memory error, Hu et al. \cite{hu2016data} first located the function in the target program where the memory error occurred. Then, they identified the gadget dispatchers containing the vulnerable code and collected data-oriented gadgets for simulating MINDOP operations.  
Second step: The expected malicious MINDOP program was taken as input, where each MINDOP operation was simulated by data-oriented gadgets with equivalent functionality. The appropriate gadgets were selected according to priority.  
Third step: Once the gadgets implementing the desired functionality were obtained, the next step was to verify the correctness of the stitching. This involved constructing concrete input to the program, triggering memory errors, and activating gadgets. If the attack was not successful, they rolled back to the second step, selected different gadgets, and attempted the stitching process again.

Ispoglou et al. \cite{ispoglou2018block} introduced a new Turing-complete attack vector known as the Block-Oriented Programming (BOP) attack. The attack payload is generated by BOPC, a tool designed to automatically construct non-control-data attacks. The BOP attack is a variant of the Data-Oriented Programming (DOP) attack. Similar to the DOP attack, the BOP attack consists of a sequence of functional blocks, which simulate each statement in the SPL payload, and dispatcher blocks, which stitch together these functional blocks.
The BOPC tool requires three inputs:
1) The exploit payload written in SPL, which specifies the attacker's intent.
2) The target program, which contains an arbitrary memory write vulnerability.
3) The entry point, which is the first instruction in the binary where the payload execution should begin. Identifying the entry point is part of the vulnerability discovery process, and it is located immediately after all Arbitrary Memory Write Primitives (AWPs) have been executed.

The principle of BOPC tool attack construction is as follows:
1) SPL Payload Generation: The attacker uses the SPL language to generate an SPL payload based on the intended malicious objective.
2) Selecting Functional Blocks: The binary front-end of BOPC utilizes angr \cite{shoshitaishvili2016sok} to decompose the target binary into basic blocks, which collectively represent the control flow graph (CFG) of the target program. Symbolic execution \cite{king1976symbolic} is employed to abstract the constraints of each basic block, capturing their impact on the program state, such as register and memory modifications, system calls, or conditional jumps at the end of the block. 
A block qualifies as a candidate if it matches the semantics of at least one SPL statement. Based on this criterion, BOPC identifies a set of candidate blocks and constructs two key mapping graphs:  
The Register Mapping Graph ($R_G$), which associates virtual registers with hardware registers.  
The Variable Mapping Graph ($V_G$), which links payload variables to underlying memory addresses.  
BOPC determines the concrete registers and memory addresses that should be preserved for each statement by searching for a maximum bipartite match in $R_G$ and $V_G$, respectively. By integrating this result with the set of candidate blocks, BOPC identifies which of these blocks can serve as functional blocks.  
After determining the set of functional blocks ($FB$) for each SPL statement, BOPC constructs the delta graph ($\delta_G$), a multipartite directed graph where each set of nodes corresponds to functional blocks representing a single payload statement. An edge between two functional blocks signifies the minimum number of basic blocks required to transition from one functional block to another. The minimum induced subgraph is then extracted from this graph, ensuring that each IR statement is mapped to a single functional block.
3) Searching for Dispatcher Blocks: Based on the concrete functional blocks obtained in the previous step, BOPC employs concolic execution to gather the necessary constraints for execution along each vertex of the minimum induced subgraph, starting from the predefined entry point. This process identifies the required basic blocks (i.e., dispatcher blocks) and the constraints necessary to stitch functional blocks together.
4) Synthesizing Exploits: The concrete constraints derived in the previous step are enforced by writing them into the target program using arbitrary memory write primitives (AWPs).

\subsection{Classification of Non-Control-Data Attacks}
In the evolutionary process of non-control-data attacks, their practical harm was first recognized, and they were later proven to be Turing complete. From the perspective of modifying non-control-data, these attacks can be classified into two main categories: security-sensitive function attacks and data-oriented programming (DOP) attacks \cite{hu2016data,ispoglou2018block}.
The first category, security-sensitive function attacks, includes data-oriented exploits (DOE) \cite{hu2015automatic} and control-flow bending (CFB) attacks \cite{carlini2015control}. These attacks manipulate security-critical data to invoke security-sensitive functions, leading to malicious outcomes such as information disclosure or privilege escalation. For example, corrupting parameters of system calls like setuid() can elevate privileges.
The second category, DOP attacks, modifies non-control-data to influence the control flow of a program, thereby enabling the execution of gadgets. Compared to security-sensitive function attacks, which are constrained by many limitations and have restricted expressive power, DOP attacks are more versatile and, in theory, can be used to achieve arbitrary malicious objectives.

The BOP attack is considered a variant of the DOP attack for the following reasons:
Both attacks are Turing-complete non-control-data attacks that modify non-control-data to influence the control flow of a program. Similar to DOP attacks, BOP attacks leverage data modifications to execute a sequence of computations without hijacking control flow explicitly.
The construction process of a BOP attack resembles that of a DOP attack. In a BOP attack, functional blocks are used to simulate the statements in the SPL payload, whereas in a DOP attack, data-oriented gadgets simulate statements in the MINDOP language. Additionally, both attacks require the identification of a dispatcher mechanism to chain the functional blocks or gadgets together.
The key distinction between the two attacks lies in the stitching phase. In a DOP attack, the stitching process is performed manually, requiring human effort to link gadgets appropriately. In contrast, a BOP attack achieves full automation, with the stitching process handled by the BOPC tool, making the attack construction more efficient and scalable.

\section{The Defense Method of Non-control-data Attacks}
Based on the evolution and classification of non-control-data attacks discussed in Section II, we can construct a general execution model for such attacks. This model serves as a complementary graph to the one presented by Szekeres et al. \cite{szekeres2013sok}.
As illustrated in Figure 3, all types of non-control-data attacks rely on the presence of at least one exploitable memory vulnerability, which enables the attacker to trigger a memory error, causing a pointer to go out of bounds or become dangling. By leveraging such an erroneous pointer for read/write or deallocation operations, an attacker can modify a non-control-data variable to a desired value. The corrupted data variable can then be used in two ways:
Guiding the control flow of the target program to execute available gadgets.
Manipulating security-sensitive functions, such as altering function parameters or decision-making variables.
Different attack types exploit these principles in distinct ways. DOP attacks corrupt data pointers to stitch data-oriented gadgets, while DOE attacks modify data pointers to achieve multi-edge stitching. One common technique is to exploit buffer overflows to overwrite data pointers, which represents an example of the backward loop in Figure 3.
For instance, consider a buffer overflow that causes an array pointer to go out of bounds in the first iteration. In the second iteration, this out-of-bounds pointer is exploited (Step 3) to corrupt a nearby data pointer in memory. When the corrupted data pointer is later dereferenced (Step 2), an attacker-controlled bogus pointer is used, effectively redirecting program execution to an attacker-chosen location.

According to the implementation steps of non-control-data attacks, the defense strategies against such attacks can be categorized into three types:
1) Memory safety. If an attacker is prevented from triggering a memory error, all memory corruption attacks, including non-control-data attacks, become infeasible. This defense strategy targets Step 1 and Step 2 in the non-control-data attack graph.
2) Data confidentiality. If the attacker is unable to determine the actual value of a data variable, they cannot modify it to a specific value required for the attack. This approach mitigates vulnerabilities at Step 4 in the non-control-data attack graph.
3) Data integrity. If an attacker is prevented from modifying a data variable or leveraging a modified variable, the non-control-data attack cannot be successfully executed. This defense mechanism addresses Step 3 and Step 5 in the non-control-data attack graph.
In the following sections, we will provide a detailed discussion of the various methods employed to defend against non-control-data attacks.

\begin{figure}[!t]
	\centering
	\includegraphics[width=3.5in]{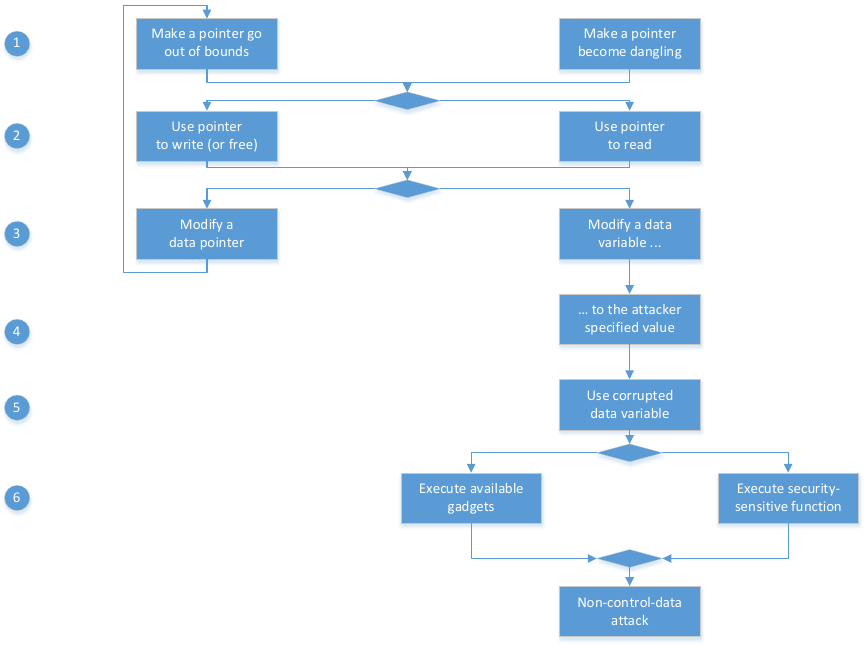}
	\caption{Non-control-data attack model.}
	\label{fig2-3}
\end{figure}

\subsection{Memory Safety}
The fundamental prerequisite for executing memory corruption attacks is the presence of at least one memory vulnerability. If complete memory safety can be guaranteed, all memory corruption attacks, including non-control-data attacks, can be effectively prevented. Complete memory safety consists of two aspects: spatial memory safety and temporal memory safety. Spatial memory safety defends against spatial errors, which originate from out-of-bounds pointer access. As shown in Figure 3, when a pointer exceeds the boundary of its referenced object (Step 1), it becomes invalid, and dereferencing this pointer (Step 2) leads to a spatial error. To enforce spatial memory safety, two approaches can be adopted: object-based approaches, which ensures that a pointer remains within a valid memory range when it is operated on, and pointer-based approaches, which checks the validity of a pointer at the moment it is dereferenced. Temporal memory safety, on the other hand, defends against temporal errors arising from the use of dangling pointers. When an object is deallocated (Step 1), any pointer referencing it becomes dangling, and the memory may later be reallocated for a different purpose. If the dangling pointer is dereferenced (Step 2), a temporal error occurs. To enforce temporal memory safety, two strategies can be implemented: disable dangling pointers, which ensures that pointers to deallocated objects cannot be used improperly, and  pointer dereference validation, which identifies and mitigates the use of stale pointers before they cause security vulnerabilities. Ultimately, the core principle of memory safety is to ensure that every memory access remains valid both spatially and temporally.

\subsubsection{Spatial memory safety}
Spatial Memory Safety: Numerous methods have been proposed to prevent spatial violations. According to the above analysis, these methods can be categorized into object-based approaches (which perform bounds checking on pointer operations) and pointer-based approaches (which validate pointers at the time of dereference).

1) Pointer-based approaches: This approach maintains base and bound information for each pointer. Early studies, such as SafeC \cite{austin1994efficient}, CCured \cite{necula2005ccured}, and Cyclone \cite{jim2002cyclone}, use a fat pointer representation to replace pointers in the program. This representation stores not only the pointer value but also the addresses of the upper and lower bounds of the object it points to. When the pointer is dereferenced, the system checks whether it is within the valid range. However, these approaches modify the pointer representation, thereby affecting the memory layout and causing incompatibility with existing binaries. Additionally, some approaches, such as CCured, even require modifications to the source code.  
SoftBound \cite{nagarakatte2009softbound} addresses these issues by not altering the pointer representation—that is, it does not bind metadata (i.e., base and bound information) directly to the pointer. Instead, it stores the metadata separately for each pointer in a hash table or shadow memory. SoftBound's method of decoupling pointers from metadata partially alleviates the binary compatibility issue. However, storing metadata separately increases memory overhead, and its performance overhead can reach up to 79\%.  
Intel MPX \cite{oleksenko2018intel} provides a hardware implementation of the SoftBound method supported by the ISA (Instruction Set Architecture), but its performance overhead remains high.

Delta Pointers \cite{kroes2018delta} addresses this problem by storing metadata in the high bits of 64-bit pointers. The lower 32 bits represent the virtual address, while the upper 32 bits store the tag bit. Within these upper 32 bits, the most significant bit (MSB) serves as the overflow bit, while the remaining 31 bits form the delta tag, which encodes the distance from the current pointer to the end of the object. During pointer arithmetic, any operation on the address is also applied to the delta tag. If the pointer goes out of bounds, the carry generated by the delta tag sets the overflow bit to 1. When the pointer is dereferenced, a bitwise AND operation is applied to mask out the delta tag while preserving the overflow bit, converting the tagged pointer into a regular untagged pointer while keeping the overflow bit intact. Consequently, if the most significant bit is set, the memory management unit (MMU) generates a fault. By offloading bounds checking to the MMU, this approach achieves a performance overhead of only 35\%. However, since half of the pointer's bits are used for metadata encoding, this method restricts the addressable virtual address space and reduces Address Space Layout Randomization (ASLR) entropy.

Pointer-based approaches can achieve complete spatial memory safety but suffer from binary incompatibility issues. Fat pointers and tagged pointers modify the pointer representation, which breaks binary compatibility. Although SoftBound does not change the pointer representation, when an unprotected library modifies a protected pointer, its metadata cannot be updated in a timely manner, inevitably leading to binary compatibility issues.

2) Object-based approaches: This approach tracks metadata information for each object to solve binary compatibility issues. Jones and Kelly \cite{jones1997backwards} proposed maintaining a splay tree (a data structure that maps a pointer to the object it points to), which contains the bounds information for each allocated object, including static objects, dynamically allocated objects, and stack objects. During pointer operations, lookups are performed in this splay tree to check whether the newly calculated address remains within bounds. This approach focuses on pointer arithmetic and is based on the principle that an address (pointer) computed from an in-bounds pointer must still point to the same object as the original pointer. To support legal out-of-bounds pointers, all allocated objects (except for function arguments) are padded with one extra byte. However, temporarily out-of-bounds pointers that extend more than one byte beyond the object’s boundaries can still trigger false alarms.

CRED \cite{ruwase2004practical} addresses this issue by creating a unique out-of-bounds object in the heap for each out-of-bounds address value and associating the value with the address of that out-of-bounds object. The out-of-bounds object contains the out-of-bounds address value and the address of the reference object to which the value refers. Its address is stored in a hash table. When pointer arithmetic is performed, if the pointer is an out-of-bounds pointer, the hash table is queried to find the address of the corresponding out-of-bounds object, thereby determining the reference object and the actual pointer value. The performance overhead of CRED is around 2x, whereas the performance overhead of the Jones \& Kelly (J\&K) approach can reach up to 11-12x.

To reduce performance overhead, Baggy Bounds Checking (BBC) \cite{akritidis2009baggy}, a technique based on the binary buddy allocator, constrains the size and alignment of allocated objects to powers of two by padding them. These constraints allow the bounds information (i.e., size and base) to be represented in a single byte and stored in a contiguous array. BBC partitions memory into fixed-size slots, each 16 bytes in size, and creates an entry for each slot in the bounds table. If an allocated object spans multiple slots, a single byte of bounds information is written to multiple slots when updating the bounds table. Finally, by setting the most significant bit of an out-of-bounds pointer, the MMU generates an error once the pointer is dereferenced. By checking allocation bounds instead of object bounds, BBC reduces the average runtime overhead to 60\%.

As mentioned earlier, object-based approaches solve the binary compatibility problem because the bounds metadata information can be updated by tracing the malloc family functions that allocate and deallocate objects from an unprotected library. However, the main disadvantage of these approaches is that they cannot provide complete spatial memory safety. For example, they do not offer effective protection for arrays of structures.

\subsubsection{Temporal memory safety}
Bounds checking can enforce spatial memory safety but cannot prevent temporal memory safety vulnerabilities such as use-after-free and double-free. According to the two steps of temporal memory violations mentioned above, it can be concluded that preventing either step from occurring is sufficient to defend against temporal memory safety vulnerabilities.

1) disable dangling pointers: This approach focuses on the first stage of defense and primarily employs two strategies to invalidate dangling pointers. The first strategy is to prevent unsafe memory reuse by using a memory allocator to ensure that the memory region pointed to by the dangling pointer (i.e., the freed memory space) is not reallocated. Although this method can defend against use-after-free attacks, it incurs high memory overhead. MarkUs \cite{ainsworth2020markus} is a memory allocator that prevents use-after-free attacks by delaying the reuse of manually freed objects until it is determined that no dangling pointers reference the freed memory region. While this method does not prevent dereferencing through dangling pointers, it ensures that the memory pointed to by a dangling pointer is never reused, thus achieving protection against use-after-free exploits. However, it is evident that allocator-based approaches cannot prevent the unsafe reuse of stack-allocated objects.

Another strategy is to prevent the creation of dangling pointers. The core idea of this strategy is either to monitor dangling pointers as soon as they are created or to invalidate pointers that reference freed memory areas. Undangle \cite{caballero2012undangle} uses taint tracking to identify all dangling pointers to a memory region after it has been deallocated. When the lifetime of these dangling pointers exceeds a certain threshold (set to 5000 instructions in the paper), they are considered unsafe dangling pointers, and this information is collected for vulnerability analysis. However, since this approach relies on high-overhead taint analysis, it has significant runtime overhead. DangNull \cite{lee2015preventing} employs a red-black tree data structure to record the address ranges of objects (obtained by intercepting all memory allocation and deallocation functions) and the relationships between objects (captured by tracking instrumented pointer assignment instructions in the target program). Once an object is freed, DangNull invalidates all pointers to that object based on the collected object relationships.

While pointer invalidation methods do not incur higher memory overhead than safe memory allocator methods, they usually introduce high runtime overhead because they need to monitor memory allocations and maintain relevant metadata. Moreover, both methods are limited to protecting dangling pointers to the heap and fail to detect all temporal errors.

2) pointer dereference validation: This method focuses on the second stage (i.e., pointer dereference or free) and prevents temporal memory violations by validating each memory access involving a pointer dereference. CETS \cite{nagarakatte2010cets} provides temporal memory safety by maintaining a unique identifier for each object and associating it with pointers to that object in a disjoint memory space. It then checks whether the object referenced by the pointer is still allocated at the time of dereference. However, storing metadata in disjoint memory not only requires spatial memory mechanisms to ensure the security of this information but also increases the number of memory accesses, resulting in high runtime overhead, averaging up to 48\%. PTAuth \cite{farkhani2021ptauth} mitigates this issue by hooking allocation functions, randomly generating a unique ID for each object, and leveraging the Pointer Authentication Code (PAC) mechanism provided by ARM to compute an encrypted authentication code (AC) based on the ID and the base address of the object pointed to by the pointer. The object metadata ID is stored at the beginning of each object, while the pointer metadata AC is stored in the unused bits of the pointer. When an object is deallocated, its ID is set to zero, making it invalid. Upon pointer dereference, the AC is recalculated and compared with the AC stored in the pointer. This inline metadata scheme employs random generation and encryption to increase the difficulty of tampering with spatial memory violation metadata while also reducing runtime and memory overhead to 26\% and 2\%, respectively. Unfortunately, this defense scheme relies on the specific hardware structure of the ARM platform and thus cannot be directly applied to other platforms.

To achieve complete memory safety, both spatial safety and temporal safety must be ensured; however, this leads to significant runtime overhead. For example, CETS and SoftBound have been formalized to guarantee temporal safety and spatial safety, respectively. Deploying CETS and SoftBound together can result in an average execution time overhead of up to 116\% \cite{nagarakatte2015everything}. BOGO \cite{zhang2019bogo}, a complete memory safety solution based on Intel MPX, incurs a runtime overhead of 60\%.

Due to the inherent flaws of the C/C++ language, such as the lack of bounds checking and manual memory management, as well as the aforementioned complete memory safety enforcement schemes, which are too costly at runtime to be practically deployed, memory vulnerabilities are expected to remain widespread for a long time. The strategy of maintaining data confidentiality is to encrypt memory data so that even if an attacker exploits memory vulnerabilities to modify the data, they cannot modify it to the expected value. A typical representative method of this strategy is Data Space Randomization (DSR). Compared to Address Space Randomization (ASR), which primarily targets control-flow hijacking attacks, DSR can defend against all memory corruption attacks.

The DSR (Data Space Randomization) \cite{bhatkar2008data,cadar2008data} scheme is inspired by PointGuard \cite{cowan2003pointguard}. PointGuard protects pointers from tampering by encrypting them while they are stored in memory. The system randomly assigns a key to each process, which is stored on a read-only page, and encrypts pointers using pointer value XOR key. When the pointer is dereferenced, i.e., loaded into registers, it is decrypted by XOR’ing it again with the same key. However, PointGuard only protects pointer data and cannot protect non-pointer data, making it vulnerable to non-control-data attacks. Furthermore, since all pointers are encrypted with a single key, once an encrypted pointer is leaked, the key can be easily recovered by an attacker.
To address these shortcomings, DSR \cite{bhatkar2008data} partitions all variables into different equivalence classes based on the points-to graph obtained through pointer analysis. If multiple pointers can be stored or loaded through the same pointer dereference, they are considered to belong to the same equivalence class. DSR randomly assigns a key to each equivalence class and encrypts variables in memory by XOR’ing their values with the corresponding equivalence class key, decrypting them with the same key before use.
However, this scheme employs the Steensgaard algorithm, which is a context-insensitive pointer analysis method, to improve runtime performance. As a result, many unrelated variables may be mistakenly classified into the same equivalence class, leaving significant room for exploitation by attackers.

HARD \cite{belleville2018hardware} is a hardware-assisted implementation of DSR (Data Space Randomization), which provides stronger security than pure software-based DSR by using context-sensitive pointer analysis. However, this approach introduces two additional caches to accelerate encryption operations and protect encryption keys, making it difficult to deploy in practice.
Previous DSR schemes have all used simple XOR-based encryption. Once the encrypted value of known data is compromised, the encryption key can be easily cracked. Palit et al. \cite{palit2019mitigating} use pointer analysis to determine all pointers that may reference developer-marked sensitive data and value flow analysis to track all variables and objects to which the marked sensitive data may propagate. These analyses are performed throughout the entire program.
Based on this, all marked sensitive data and its related data are identified. Then, memory reads and writes to these data are decrypted and encrypted by instrumenting AES encryption operations. Although this approach provides stronger AES encryption, due to the high cost of encryption and decryption, it can only encrypt a limited amount of security-critical data. Additionally, these sensitive data must be manually marked by developers, making the approach impractical for deployment in real-world applications.

The static DSR (Data Space Randomization) scheme described earlier can be bypassed by key reuse attacks, as pointed out by Palit et al.
To address this issue, Palit et al. proposed a dynamic DSR approach, known as CoDaRR \cite{rajasekaran2020codarr}, which periodically or on-demand re-randomizes the masks during program execution and updates the program's memory representation accordingly. Specifically, it updates all masks in the code, registers, and stack, and re-encrypts all global, stack, and heap data that were previously encrypted with the old masks.
However, this scheme fails to accurately distinguish function pointers from constants when updating function pointers, which may cause the program to crash at runtime due to the use of stale encrypted function pointer values. 

\subsection{Data Integrity Protection}
In addition to using the DSR method to prevent an attacker from modifying data as expected (corresponding to Step 4 in Figure 3), a defense can also prevent an attacker from modifying data by enforcing data integrity (corresponding to Step 3 in Figure 3). Alternatively, by enforcing data flow integrity, an attacker can be detected when using the modified data, even if they are able to modify it.

\subsubsection{Data integrity}
When an attacker writes illegal data, data integrity schemes can detect and prevent unauthorized modifications. Yong et al. \cite{yong2003protecting} proposed a security enforcement tool for C that protects against writes via unchecked pointer dereferences.
The tool uses static analysis to identify unsafe pointers (i.e., pointers that might go out of bounds due to pointer arithmetic or non-NULL pointers that might become dangling) and tracks the memory locations these pointers can legitimately reference (tracked locations). By instrumenting the program, all locations in the mirror region (a memory area where each bit represents a byte in program memory) are initially marked as inappropriate. As the program runs, these tracked locations are marked appropriate upon allocation and inappropriate upon deallocation.
Each time a write operation occurs via an unsafe pointer dereference, the system checks whether the target location is marked as appropriate. If the tag is inappropriate, an error is reported, and the program is terminated.
This approach effectively protects against most attacks, such as stack smashing and other exploits that illegally modify data via unsafe pointers, by verifying at runtime whether the memory location being accessed is a valid target for an unsafe pointer. However, this method simply classifies all memory locations as either appropriate or inappropriate. If adjacent memory locations are accessed by different unsafe pointers or the same unsafe pointer, the protection mechanism may fail.

To address the shortcomings of the previous technique—where all object locations are classified into only two categories (whether they are suitable for writing via unsafe pointer dereferences) and represented using just 1 bit, making the approach too coarse-grained and easily bypassed—Write Integrity Testing (WIT) \cite{akritidis2008preventing} improves this by using 8 bits, allowing it to categorize all objects into 256 types.
The core principle of WIT is leveraging points-to analysis to classify all objects into multiple sets based on the write instructions present in the program. Each set consists of objects that can be written by the same write instruction, and WIT assigns a unique color to each write instruction and its corresponding set of objects.
At runtime, WIT maintains a color table, which maps all memory locations to their corresponding colors. Each entry in this table is an 8-bit color identifier, representing 8 bytes of memory, and is dynamically updated as objects are allocated and deallocated.
Before executing a write instruction, WIT verifies that the color of the memory location being written (from the color table) matches the color assigned to the instruction. Additionally, WIT inserts small guard areas, similar to Canary values, between objects to prevent sequential overflows and underflows.
Furthermore, WIT enforces Control-Flow Integrity (CFI) by ensuring that indirect call instructions are assigned the same color as the functions they are allowed to invoke.
WIT incurs an average space overhead of 13\% and an average runtime overhead of 7\%. However, it has the following limitations:
The 8-bit differentiation may not be sufficient for very large programs, making it less effective in certain scenarios.
Both Yong’s method and WIT only enforce write integrity—they do not protect reads, meaning information leakage remains a potential vulnerability.

\subsubsection{Data-flow integrity} Data-Flow Integrity (DFI) \cite{castro2006securing} uses static analysis to compute a Data-Flow Graph (DFG) and instruments the program to ensure that the runtime data flow conforms to the DFG.
The DFG is computed based on Reaching Definitions Analysis. For each instruction that loads (reads) data from memory, DFI computes the set of instructions that may store (write) the data. DFI instruments reads to check whether the last instruction that wrote the data belongs to the set computed by static analysis.
To record the last instruction that wrote to a memory location at runtime, DFI maintains a table that stores the identifier of the last instruction that wrote to each memory location and updates the table before every write.
To prevent an attacker from tampering with the table, DFI also performs integrity checks before each write instruction.
However, this approach incurs an average runtime overhead ranging from 43\% to 104\%, making it difficult to deploy in real-world applications.

Hardware-Assisted Data-Flow Isolation (HDFI) \cite{song2016hdfi} enforces data-flow isolation by virtually extending each memory unit with an additional tag. It introduces new instructions such as sdset1, which sets the tag to 1 when storing sensitive data, and ldchk1, which checks whether the tag is 1 when loading the data. Since regular store instructions do not set the tag, any attempt to load sensitive data with ldchk1 after being stored using a regular store instruction triggers a tag mismatch and reports an error. Although HDFI has a low performance overhead (<2\%), its single-bit granularity is too coarse, making it unable to detect data-flow deviations with the same tag value. Tagged Memory Supported Data-Flow Integrity (TMDFI) \cite{liu2018tmdfi} extends memory tag bits to 8 bits to store data-flow identifiers, providing finer granularity but still limited to only 256 unique identifiers, which is insufficient for large programs. TMDFI incurs a significant performance overhead of 39\%. Feng et al. \cite{feng2021toward} identified that the main performance bottleneck of software-based DFI is the DFI check instructions, and to mitigate this, they offloaded DFI checks to a Processing-In-Memory (PIM) processor \cite{azarkhish2016design}, significantly reducing performance overhead while maintaining full DFI enforcement. However, the complexity of this approach makes it nearly impossible to deploy in real-world applications. KPDFI \cite{nie2023kpdfi} narrows the scope of protected data to three types related to memory corruption attacks—Exploited, Unsafe, and Input-related—reducing the overall overhead to just 9.53\%. However, its reliance on static analysis to define multiple data types makes it challenging to implement and may result in a high false negative rate.

\section{Defense Methods Against DOP Attacks In Recent Years}
Among the general defense methods against memory corruption attacks mentioned above, software-based solutions that do not modify hardware structures often incur excessively high performance overhead. In recent years, some methods have been proposed to defend against non-control-data attacks while reducing performance overhead. These methods can be mainly classified into two categories: 1) reducing the scope of protected data to lower runtime overhead; 2) focusing defense on the stack, as non-control-data attacks frequently occur there.

\subsection{Reduce the data that needs to be protected}
Datashield \cite{carr2017datashield} divides program data into sensitive data and non-sensitive data and places them in two separate memory regions. The sensitive data is annotated by the programmer. Datashield performs precise bounds check only on pointers to sensitive data and do a coarse bounds check on pointers to non-sensitive data to ensure that they point anywhere but the sensitive data area. Although Datashield only does coarse bounds check on a large number of pointers to non-sensitive data to reduce the performance overhead, it requires a programmer to determine by experience which types of data are sensitive.

TRUVIN \cite{geden2020truvin} employs a mandatory value-based integrity method similar to OAT to detect whether a program has been attacked by a non-control-data attack. The difference is that critical variables do not need to be marked by the programmer but are automatically identified by the protection tool. Its principle is to classify all variables into critical variables and non-critical variables based on the trustworthiness of their value sources. Trustworthiness has two aspects: trust source (e.g., programmer-defined values) and trust propagation (e.g., the value of a variable is copied from another trusted value). By analyzing the trust sources and trust propagation of all variable values in the program, critical variables can be automatically identified. Since TRUVIN only checks critical variables, its average performance overhead is 28\%. However, trust sources are not easy to pinpoint accurately, which limits the practical applicability of this method.

\subsection{Stack-based non-control-data defense methods}
Smokestack \cite{aga2019smokestack} employs a runtime stack layout randomization method to defend against DOP attacks. The principle is to fully permute the locations of all variables on a function's stack. For example, if there are four variables on the stack, then the fully permuted stack layout has a total of 4! (=24) different possibilities. At runtime, one of these stack layouts is selected based on a generated true random number. Additionally, to prevent attackers from bypassing this protection method, a label is set at the prologue of each function and checked at the epilogue to ensure a match. Since a different stack layout is randomly selected each time a function is called, a DOP attacker cannot determine the relative positions of objects on the stack, making it impossible to construct an effective attack. However, this method of randomizing the stack layout at runtime inevitably results in excessive performance overhead, especially for large-scale programs.

SafeStack+ \cite{lin2017safestack} is a defense method against DOP attacks based on SafeStack \cite{kuznetzov2018code} stack technology. SafeStack divides the regular stack into two parts: a safe stack and an unsafe stack. All objects that are proven to be safe are placed on the safe stack, while objects that cannot be proven safe are placed on the unsafe stack. Objects on the unsafe stack cannot corrupt any object on the safe stack through methods such as buffer overflows, effectively defending against control-flow hijacking attacks. However, objects on the unsafe stack can still overwrite each other and may be vulnerable to non-control-data attacks. SafeStack+ uses the def-use analysis method to identify all unsafe stack variables that may affect the execution of conditional branch statements and inserts random-length and random-value canaries around these critical variables. At runtime, when evaluating branch conditions, the program checks whether the canary values have changed to detect potential DOP attacks. However, this defense strategy only considers variables that affect the execution of conditional branch statements as critical variables, meaning it can only prevent DOP attacks but cannot protect against security-sensitive function attacks.

\section{Conclusions and future directions}
With the increasing number of defense strategies against control-flow hijacking attacks, academia and industry are also paying growing attention to non-control-data attacks. In this review paper, we first categorize non-control-data attacks into two phases based on their evolution and establish a model of non-control-data attacks according to the steps required to execute them. Subsequently, we identify three strategies for defending against non-control-data attacks based on this model. Then, we summarize and analyze recent methods specifically designed to defend against Data-Oriented Programming (DOP) attacks. Although several defense strategies against non-control-data attacks have been proposed, they are difficult to deploy commercially due to high execution overhead, the need for hardware modifications, or compatibility issues. To address these challenges, we propose several interesting future research directions in this field.

First, a more general DOP automation exploitation framework. The DOP exploitation framework proposed by Hu et al. is semi-automated, while the automated framework BOP proposed by Ispoglou has achieved a certain level of automation. However, these methods remain prototype implementations with limited functionality.

Second, complete memory security. Memory bugs are the root cause of all memory corruption attacks, and deterministic methods such as WIT and DFI are merely approximate solutions for enforcing memory safety. Existing complete memory security mechanisms suffer from high execution overhead. Therefore, research in this direction, especially on temporal memory security, is worthy of attention.

Finally, using machine learning methods to detect non-control-data attacks. The key challenges of this defense approach are how to obtain sufficient training and testing datasets and how to address the issue of excessively high false positives.

\newpage

\begin{IEEEbiography}[{\includegraphics[width=1in,height=1.25in,clip,keepaspectratio]{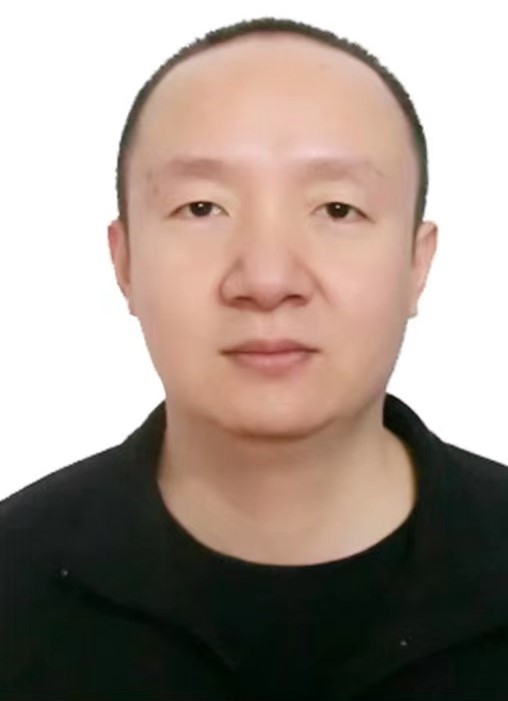}}]{Lei Chong}
	received his BS degree in automation from Wuhan Textile University, Wuhan, China, in 2005,
	and MS degree in control theory and control engineering from Guangdong University of Technology, 
	Guanzhou, China, in 2010. Currently, He is a lecturer at Sichuan University of Arts and Science. His research interests include system security and software security. E-mail: 20170005@sasu.edu.cn
\end{IEEEbiography}

\end{document}